\documentclass[aps,prl,showpacs,floatfix,superscriptaddress]{revtex4}
\usepackage{times,graphicx,amsmath,bbm,mathrsfs,amssymb,psfrag,color}

\newcommand{\be}{\begin{equation}}
\newcommand{\ee}{\end{equation}}
\newcommand{\bea}{\vspace{0.25cm}\begin{eqnarray}}
\newcommand{\eea}{\end{eqnarray}}

\begin{document}

\title{Experimental Test of an Event-Based Corpuscular Model Modification as an Alternative to Quantum Mechanics
 }
\author{Giorgio Brida$^{1}$, Ivo Pietro Degiovanni$^{1}$, Marco Genovese$^{1}$\footnote{E-mail: m.genovese@inrim.it}, Alan Migdall$^{2}$,
 Fabrizio Piacentini$^{1}$, Sergey V. Polyakov$^{2}$, Paolo Traina$^{1}$ }
\affiliation{1- Istituto Nazionale di Ricerca Metrologica, Strada delle Cacce 91, Torino 10135, Italy; 2-  Joint Quantum Institute and National Institute of Standard and Technology, 100~Bureau~Dr.
Stop 8410, Gaithersburg, MD 20899, U.S.A.}

\begin{abstract}  We present the first experimental test that distinguishes between an event-based corpuscular model (EBCM) [H. De Raedt et al.: J. Comput. Theor. Nanosci. {\bf 8} (2011) 1052] of the interaction of photons with matter and quantum mechanics. The test looks at the interference that results as a
single photon passes through a Mach-Zehnder interferometer [H. De Raedt et al.: J. Phys. Soc. Jpn. \textbf{74} (2005) 16]. The experimental results, obtained
with a low-noise single-photon source [G. Brida et al.: Opt. Expr. {\bf 19} (2011) 1484], agree with the predictions of standard quantum
mechanics with a reduced $\chi^2$ of 0.98 and falsify the EBCM with a reduced $\chi^2$ of greater than 20.\end{abstract}

\maketitle

\section{Introduction}

Quantum mechanics (QM) is a pillar of modern physics and  its theoretical predictions are confirmed by an abundance of very accurate experimental data. Furthermore,
the theory is successfully applied to a wide spectrum of phenomena that include such disparate areas as solid state physics, cosmology, bio-physics, and particle physics.

Nevertheless, even after nearly a century of debate, problems related to the foundations of this
theory persist \cite{genovese,ab,f}, particularly the transition from the macroscopic deterministic world described by classical mechanics (macro-objectivation) to the microscopic
probabilistic world and the concept of measurement as described by quantum mechanics. In addition, interpretations of purely quantum phenomena, such as wave-function
collapse and nonlocality, have not been definitively resolved, because either appropriate tests have not been developed or the accuracy of existing tests is insufficient.
With the development of promising quantum technologies such as quantum
information (computation, communication, etc.)\cite{QT}, quantum metrology \cite{qm}, quantum imaging \cite{qi}, etc., as well as our current reliance on quantum-mechanics-based technology, the importance of these questions goes beyond pure theoretical interest and can directly impact practical issues.

In particular, Bell's 1965 paper \cite{bell} demonstrated that Local Hidden Variable Theories (LHVT)
cannot reproduce all the results of quantum mechanics when dealing with entangled states. Since
the introduction of that paper, many experiments have attempted to discriminate between the predictions of quantum mechanics and LHVTs using Bell's test on two-particle systems \cite{genovese}. So far no experimental Bell test has simultaneously closed all of the known loopholes. This requires at a minimum a) space-like separation of the measurements on the two particles so the results of the measurement of one particle cannot influence the other measurement of the other particle and b) high enough detection efficiency so that one need not rely on an assumption that the subset of particles detected is a fair sample of the entire population of created. Without satisfying these two conditions simultaneously the results so far are not conclusive.

The existence of these loopholes means that it is still possible to construct local realistic models that are compatible with existing
experiments. Such models offer a way of avoiding nonlocality or nonrealism, which make many people uncomfortable.
As a result, experimental tests have been proposed \cite{loop} for some interesting classes of local realistic models.

One such class \cite{dr1,dr} deals with the possibility of building a model where
photons are described as particles, while wave (quantum) interference is retained through a mechanism in which the system keeps ``memory" of previous events and undergoes an adaptive evolution.  This scheme, dubbed the Event-Based Corpuscular Model (EBCM), reproduces quantum behavior on average, but differs from quantum mechanics over short measurement times and thus allows for tests that discriminate between the two. One such test\cite{dr} features a specific configuration of a Mach-Zehnder
interferometer traversed by single-photon states. We have implemented that test experimentally and show that the results agree with QM and falsify the EBCM \cite{dr1,dr}.

\section{Summary of the Theoretical Proposal}

\begin{figure}
   \begin{center}
   \includegraphics[scale=0.75]{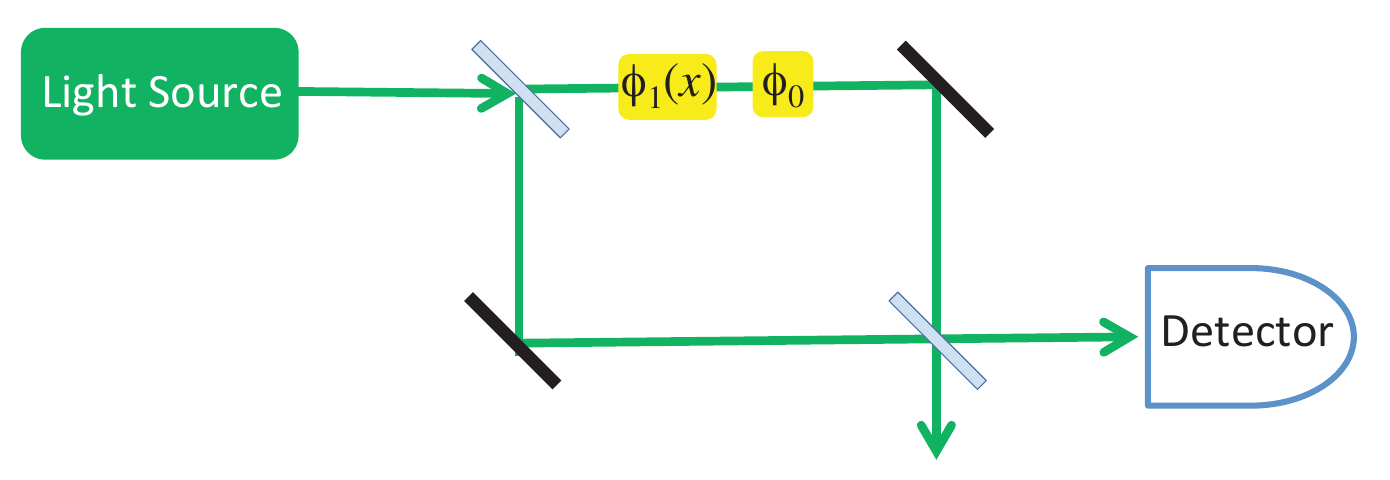}
   \end{center}
    \caption[example2]
{\label{fig:MZI} (Color online) Schematic of a device that tests a memory-based model of a beamsplitter: if a phase $\phi_1$ of a Mach-Zender interferometer is randomly set to either $0$ or $\pi/2$ for each photon that traverses the apparatus, then the output fringe would differ significantly from that predicted by quantum mechanics.}
   \end{figure}

Consider a Mach-Zehnder Interferometer (MZI), comprised of two beamsplitters connected via two optical paths, where one path's optical length can be  varied (Fig. \ref{fig:MZI}; see also Fig. 1 of Ref. \cite{dr}). A source of single photons is used as an input so only one photon is traversing the system at a time. The beamsplitters are assumed to be independent from one another, and may only acquire information about the phase difference between the two paths via the traversing photons. Each single-photon event (or experiment)
results in a detection by one of the two single photon detectors at the
outputs of the MZI. When a sufficient number of events are gathered, the expected interference
effect is observed. That is, for a series of measurements, where the phase difference varied from 0 to $2\pi$, the expected interference fringe pattern is observed, provided that enough statistics is acquired at each phase point. This result is predicted by QM and the EBCM\cite{dr1}. The ECBM achieves this because each traversal of a photon through the beamsplitter provides some information about the path length traveled by the photon, and the beamsplitter in turn gradually learns about phase difference of the two paths. Thus, while in the long run the two theories yield nearly identical results, for a small number of traversing photons the EBCM differs from QM. This is because the first photon traversal in ECBM is very different from subsequent ones. In the EBCM, the subsequent photon propagation depends on that of the previous ones due to the ``memory effect," which results in a lag in response when the incoming photons' phase changes too rapidly.

Let us now consider a situation with two controllable phases in the interferometer (fig. 1). One phase, $\phi_0$, which is varied slowly, is used to map out the familiar sinusoidal interference pattern of the interferometer.  The second phase,  $\phi_1(x)$,  is changed rapidly to switch between two values determined by a random variable $x$, $\phi_1(x=-1) = 0$  and
$\phi_1(x=1) = \pi / 2$. These values are chosen for each photon after it is created and before it enters the MZI.
While in QM the result of the
experiment is independent of the sequence of $x$ values, the ECBM\cite{dr,dr1} predicts an interference pattern that is not independent of the $x$ sequence and thus is at variance with the predictions of QM. This is because in the EBCM, the fringe observed depends on the recent history of the phases acquired by the photons traversing the MZI, the properties of the BS learning (or ``memory parameter")\cite{dr,dr1}, and on the number of photons passing through the MZI while $x$ is constant.

\begin{figure}
   \begin{center}
   \includegraphics[scale=0.5]{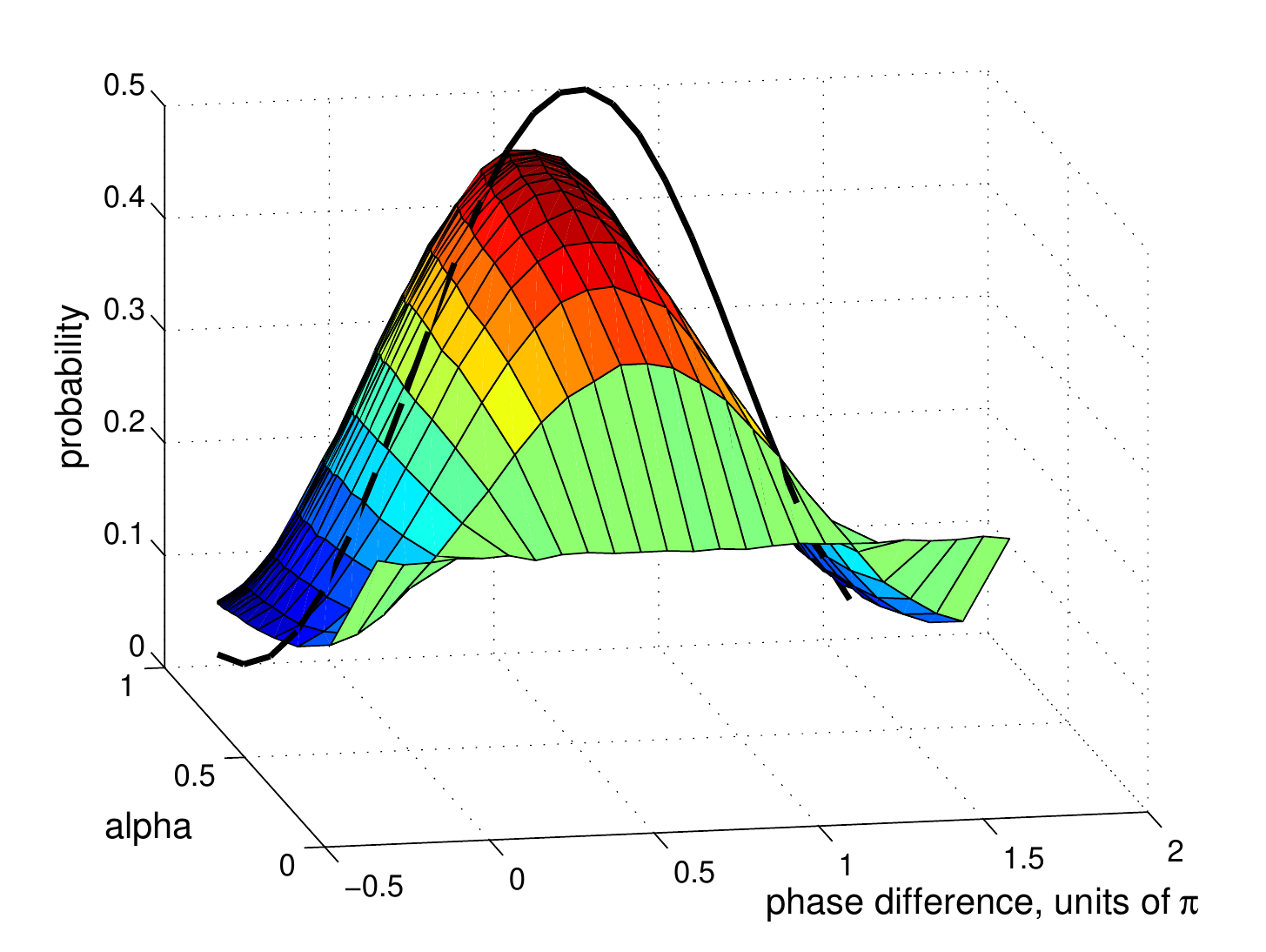}
   \end{center}
    \caption[example2]
{\label{fig:alphadependance} (Color online) Numerical simulations showing the dependence of an EBCM interference fringe for an MZI whose phase difference changes at random for each traversing photon on a memory parameter $\alpha$. Cf. quantum mechanical prediction (thick solid line). }

   \end{figure}

For example, a clear phase shift in the interference pattern is predicted ($\approx$0.57 rad, see Figs. 2,3 of Ref.\cite{dr}) between the cases when $x$ is randomly selected once for every subsequent photon as opposed to
once for every $10$ photons. Learning by a beamsplitter is characterized in Refs. \cite{dr,dr1} by a ``memory parameter" $\alpha$, which can range from 0 to 1, also play an important role. For a small $\alpha$ the model retains more information per each photon traversal, and therefore ``forgets" its previous state faster. For $\alpha\approx 1$ the situation is reversed. We numerically simulated predictions of the model for a full range of $\alpha$ values for the case where $x$ is randomly selected for every subsequent photon, Fig. \ref{fig:alphadependance}. Note that the smaller the $\alpha$, the bigger the difference between the EBCM and QM predictions. Thus expanded tests of this property can not only falsify either EBCM or QM, but also determine $\alpha$ for the EBCM.

The authors\cite{dr} point out one specific condition needed to falsify the EBCM. In particular the
switching rate must be tuned in such a way that the $x$ value is constant during the photon
traversal of the interferometer. They also mention that a limited detection efficiency does not affect the result, provided that dark counts do not contribute a significant fraction of detected events.

\section{  The Experimental Set-Up}

\begin{figure}
   \begin{center}
   \begin{tabular}{c}
   \includegraphics[scale=0.5]{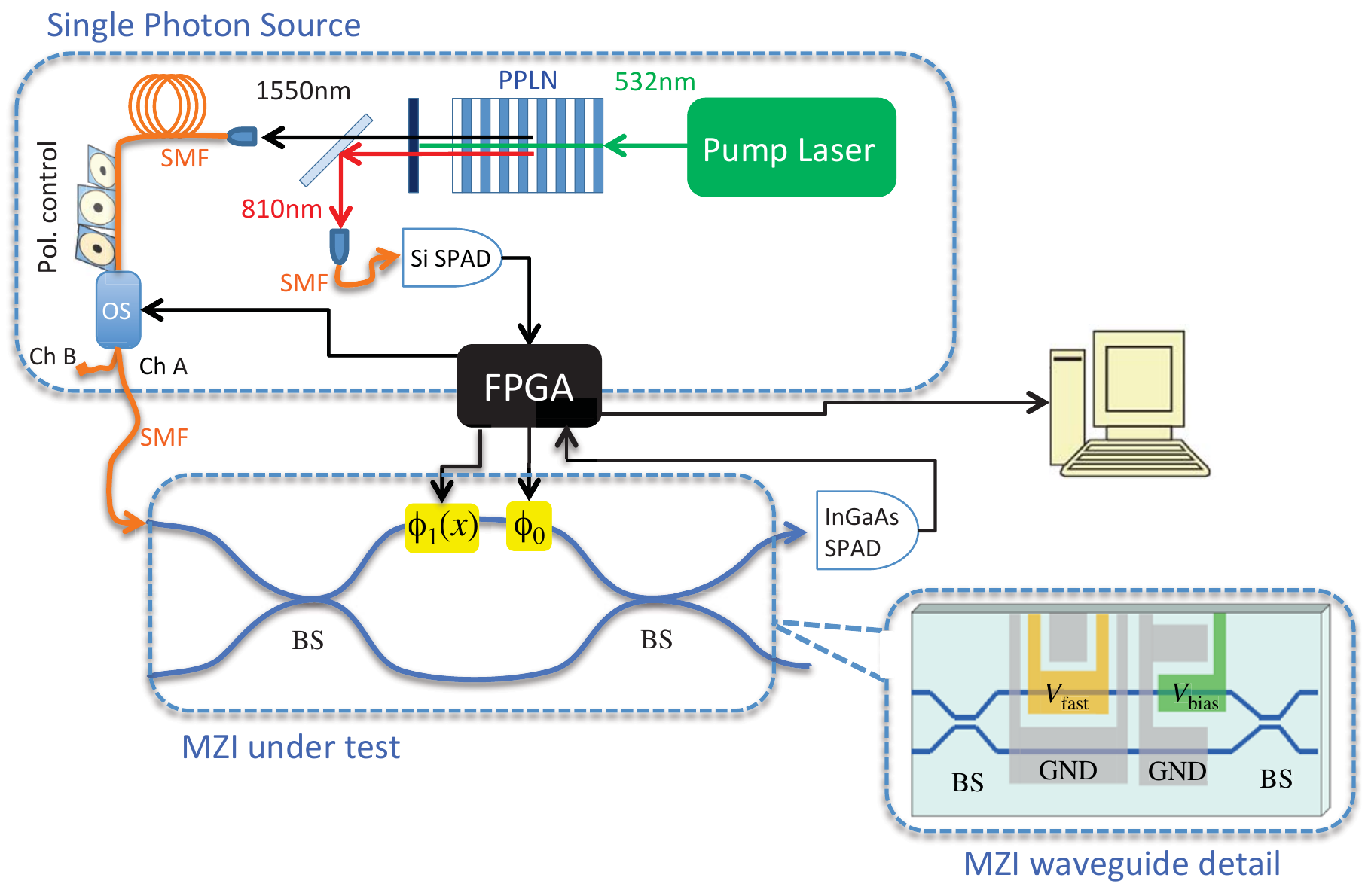}%DeRaedtpaperfigure02.pdf} %setup_Deraedt2sp_AM-pdf.pdf
   \end{tabular}
   \end{center}
   \caption[example2]
%>>>> use \label inside caption to get Fig. number with \ref{}
   { \label{fig:setup} (Color online) Experimental setup for the test of standard Quantum Mechanics versus the EBCM. The single photons are provided by a heralded photon source, with background photon flux greatly reduced by an optical switch working as a shutter opening only in the presence of a heralded single photon. The source output channel (CH A) is sent to a LiNbO$_3$ wave-guide MZI, modulated by means of a pulse generator. One output of the MZI is then sent to InGaAs detector and counting electronics.
   }
   \end{figure}

To implement this test of the EBCM, we have used a high-performance single-photon source
coupled into a MZI whose phase shift is controlled electronically. In our experimental setup
(fig.\ref{fig:setup}) pairs of photons at $\lambda_{s}=1550$ nm (signal) and $\lambda_{i}=810$ nm (idler) are generated via parametric down-conversion  in a $5\times1\times10$ mm periodically-poled Lithium Niobate (PPLN) crystal pumped by a continuous wave (cw) laser ($\lambda=532$ nm). The photon at $\lambda_{i}=810$ nm is used to herald the arrival of a $\lambda_{s}=1550$ nm photon and control an opto-electronic shutter. Thus we have a heralded single-photon source at $\lambda_{s}=1550$ nm with a low probability of emitting photons that are unheralded.
This low-noise heralded single-photon source (HSPS) using the herald controlled shuttered output principle\cite{rarity} is described in detail elsewhere \cite{NHSPS}.

The idler photon is sent through an interference filter (not shown) with a full width half maximum (FWHM) of 10
nm, then coupled into a single-mode fiber and addressed to a silicon single-photon avalanche detector (Si-SPAD). The
signal photon is sent through a 30 nm FWHM interference filter (not shown) and coupled to a single-mode optical fiber
connected to an optical switch (OS), acting as a shutter on the output of the single-photon source (OS channel A). The OS is controlled by a fast pulse generator triggered by a field
programmable gate array (FPGA). Such a setup reduces the number of background (noise) photons by opening
the HSPS output channel only when a heralded photon is expected.  The FPGA triggers the pulse
generator that opens OS channel A for a time interval $\Delta t_{\textrm{switch}}=4$ ns, then
switching to channel B for a chosen ``shuttered'' time (equal to the detector's deadtime) $t_{\textrm{dead}}=20$ $\mu$s before the
system is able to receive a new trigger by a Si-SPAD counts. The photon exiting from OS channel A has a definite polarization and its polarization is preserved in a
polarization-maintaining fiber.

The OS channel A, chosen as our low-noise HSPS output channel, is then connected to the fiber-based MZI where polarization is maintained and
matched to the polarization of the HSPS. The MZI has a LiNbO$_3$ waveguide-based phase modulator in each arm. The phase
$\phi_0$ is controlled with a bias voltage $V_{\mathrm{bias}}$ applied to the electro-optic material of the
device. The $x$ value in the MZI is controlled by another pulse generator triggered by a different
output of the same FPGA. After traversing the MZI, the photon is coupled to an InGaAs SPAD with a detection window of 50 ns.
 The output of this detector is sent to the same FPGA board for real-time data processing.

To investigate the theoretical predictions of EBCM, we mapped the output of the MZI as a function of $\phi_0$. We did this with three different measurement acquisitions: one
with the MZI $\phi_1(x)$ with $x$ fixed at $x=1$ for every single photon, a second one with the MZI $\phi_1(x)$ with $x$ fixed at $x=-1$ and a
third one with the MZI $\phi_1(x)$ switching between $x=-1$ and $x=1$ for each detected heralding photon. The last
configuration reproduces a random procedure proposed in \cite{dr} (see fig. 5 of that paper). Because $x$ is switched for every herald, and because both the emission probability of the source and the detection efficiency of the detector are low, $x$ is effectively randomized between successful trials.
We note that, in accordance with the theoretical proposal, we made sure that the
MZI did not change its phase difference during a photon traversal.

Prior to this present experiment, we quantified the fraction of higher-order photon state emission of the our HSPS with a Hanbury-Brown Twiss setup \cite{hb}
(comprised of a 50\%-50\% fiber beam-splitter with its two outputs connected to two InGaAs-SPADs).
That test showed a second-order auto-correlation of $g^{(2)}(t=0)= 0.008 \pm 0.004$ and an output
noise fraction (defined as the ratio between the background noise (i.e. unheralded) photons and the total number of
photons\cite{NHSPS}) $ONF=(0.47\pm0.02)\%$. Those results guarantee a negligible
multi-photon component and a very low number of background photons traversing the MZI. This low-noise photon source is very important in correctly implementing the experiment as a test of ECBM, because it guarantees that there are no extra photons that traverse the MZI and provide its beamsplitters with extra phase information.

\begin{figure}
   \begin{center}
   \includegraphics[scale=0.7]{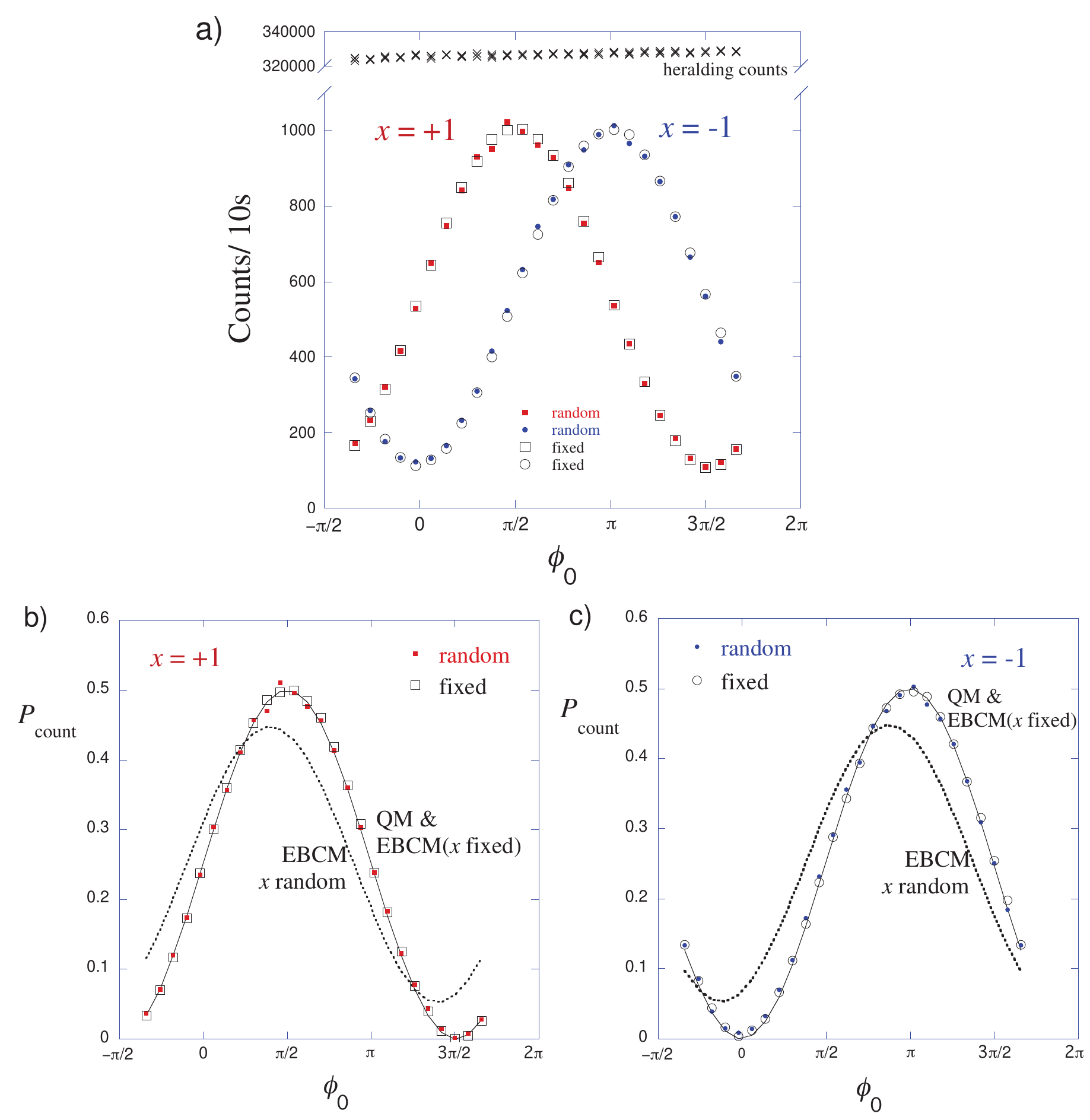}
   \end{center}
\vspace{2cm}
   \caption[example2]
   {\label{fig:data} (Color online) Experimental results. Interferometer output as $\phi_0$ is scanned slowly, while $\phi_1(x)$ is set for one of two modes: one in which $x$ is fixed (open symbols) and one in which $x$ is change randomly between detected photons (solid symbols). $x=-1$ data (circles) and $x=1$ data (squares) are indicated.
(a) Unprocessed experimental results without background subtraction, showing the dependance of the number of photoelectronic detections on phase $\phi_0$. Note the absence of a phase shift predicted by the ECBM between the fixed $x$ mode data (open points) and random $x$ mode data (filled points) for both the $x=+1$ and $x=-1$ data sets. Heralding counts are shown on top of the figure, after the
axis break, highlighting the stability of the heralding rates over the full measurement time. (b), (c)  Heralded counts corresponding to  $P_{\mathrm{count}}(x=+1)$ and $P_{\mathrm{count}}(x=-1)$ determined from the raw data. For comparison theoretical predictions are shown: the nearly identical predictions of standard quantum mechanics and the EBCM for fixed $x$ (solid curve); the  phase shifted and reduced visibility prediction of the EBCM (dashed curve). For the EBCM  $\alpha=0.99$.
 All the uncertainties (coverage factor $k=1$) are smaller than the size of data points. }

   \end{figure}

\section{  Experimental Results}

The measured output of the MZI is shown (Fig. \ref{fig:data} (a)) as a function of phase     $\phi_0$ for each
of the three measurement conditions: $\phi_1(x)$, with $x$ fixed at either $1$ or $-1$ (black dots)
and $x$ randomly switched (orange dots) corresponding to $\phi_1=0$, $\pi/2$, and random swapping between the two.  The data was acquired in the following sequence (fulfilling the requirements of
the original proposal \cite{dr}):  for each value of $\phi_0$, 10 sets of 10 second measurements were taken for $x=-1$, followed by 10 sets of measurements for $x=+1$, then 10 sets of measurements where $x$ was switched randomly. This was repeated for each value of $\phi_0$ in order until the entire full $2 \pi$ range was covered.
 It is evident that the ECBM predictions of
reduced visibility and shifted phase between the fixed $x$ and random $x$ configurations (Fig. \ref{fig:data} and Figs. 2 and 3 of ref.\cite{dr}) are not present.

\begin{figure}
   \begin{center}
   \includegraphics[scale=0.4]{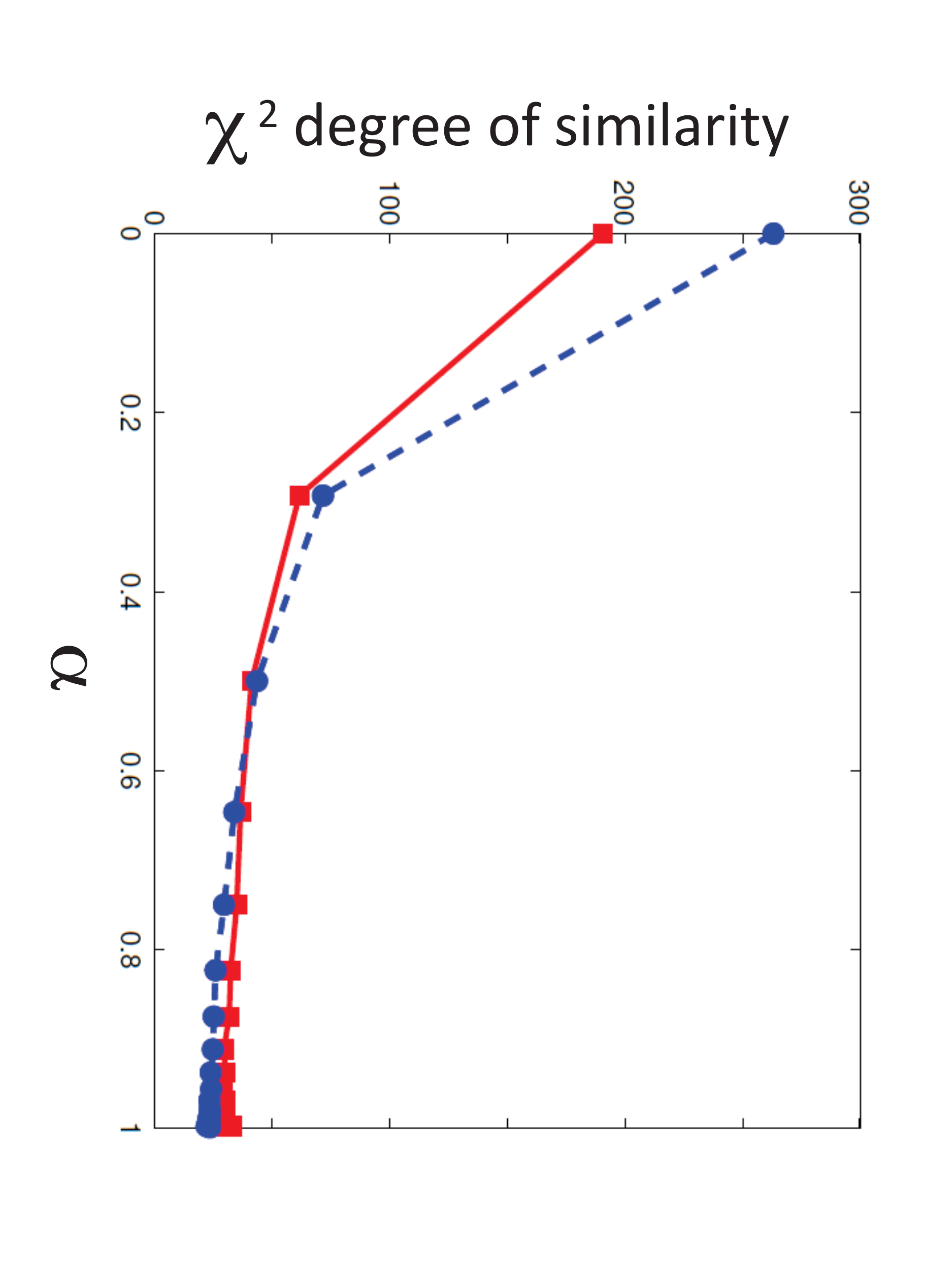}%agreement3.pdf}%fig5v2.pdf}
   \end{center}
    \caption[example2]
{\label{fig:agreement} (Color online) Agreement between an observed fringe and a theoretical prediction, showing that the observed data is inconsistent with the memory model. The two curves correspond to $x=-1$ (circles, connected with a dashed line), and $x=1$ (squares connected with a solid line).}

   \end{figure}

Fig. \ref{fig:data} (b) and (c) shows the normalized count probability $P_{\mathrm{count}}$ for $x$ fixed at either $1$ or $-1$ and for randomly switching $x$ along with the theoretical predictions given by quantum mechanics (solid curve) and the EBCM
(dashed curve) and a memory parameter of $0.99$. A $\chi^2$ test for a full range of the memory parameter shows that agreement is poor for all values of $\alpha$, Fig. \ref {fig:agreement}. As evident from the plots, the experimental results agree well with
the predictions of quantum mechanics  and significantly disagree with the predictions of EBCM, regardless of a memory parameter $\alpha$.  As evident from the plots,
the experimental results agree well with the predictions of quantum mechanics (reduced $\chi^2$
around 0.98) and significantly disagree with the predictions of EBCM, regardless of a memory
parameter $\alpha$, in fact the reduced $\chi^2$ is in this second case always greater than 20.
%Indeed the
%$\chi^2$ resulting from fits of the data to the EBCM predictions are $213$ times worse than the fit to the quantum predictions for $x=1$, and $102$ times worse for $x=-1$.

Though we subtracted dark
counts in these figures, this does not affect our conclusions, because the raw data in
Fig.\ref{fig:data} (b), (c) presents the same behavior as in Fig. \ref{fig:data} a), despite the
small reduction of the visibility due to presence of the dark counts. Obviously, also in this case the
phase shift of the fringe, predicted by EBCM, is absent.

We note that the MZI used in this study is an integrated device, so there might be some crosstalk between the phase change in the other arm of the MZI as the intended arm is varied. The manufacturer confirmed that because of the electrode design (Fig. \ref{fig:setup} MZI detail), the control field felt by the other arm can result in a phase shift in the other direction of up to $20\%$ of the controlled arm. We included this effect in the simulation of ECBM and verified that allowing a simultaneous phase change in both the arms of the interferometer may lead to a change of the EBCM fringe visibility, but such a change does not affect the principal findings of this study. Note that the EBCM of an interferometer assumes that the memory effect occurs at each of the two beamsplitters, but that they do not interact directly, rather, and all the information exchange is carried by the photons. This is exactly the case for an integrated MZI, as the two beamsplitters are distinct from the active area where dynamic phase change occurs and from each other.

\section{  Conclusions}

We have presented the first experimental test of the EBCM \cite{dr1}. This model predicts that the beamsplitters in a Mach-Zehnder interferometer accumulates and remembers
information from the photons that traverse them.To minimize uncontrolled ``learning," we used our low-noise single photon source, reducing the amount of unintended (noise) photons nearly to zero. The experimental results show an excellent agreement with the standard quantum mechanics description and conclusively falsify EBCM for a full range of allowed ``learning parameters."

\section*{  Acknowledgements}

The research leading to these results has received funding from the European Union on the basis of
Decision No. 912/2009/EC (project IND06-MIQC), by MIUR, FIRB RBFR10YQ3H and RBFR10UAUV, and by
Compagnia di San Paolo.

%%%%%%%%%%%%%%%%%%%%%%%%%%%%%%%%%%%%%%%%%%%%%%%

\end{document}